
\documentclass[twocolumn]{aastex631}

\newcommand{\halpha}{H$\alpha$}
\newcommand{\rophi}{$\rho$~Oph}
\newcommand{\vsf}{$\langle|\delta \vec v|\rangle$}

\shorttitle{Milky Way MCs Turbulence}
\shortauthors{Ha et al.}

\graphicspath{{./}{figures/}}

\begin{document}

\title{Turbulence in Milky Way Star-Forming Regions Traced by Young Stars and Gas}

\author[0000-0001-6600-2517]{Trung Ha}
\affiliation{Department of Physics, University of North Texas, Denton, TX 76203, USA}
\email{trungha@my.unt.edu}

\author[0000-0001-5262-6150]{Yuan Li}
\affiliation{Department of Physics, University of North Texas, Denton, TX 76203, USA}

\author[0000-0002-5365-1267]{Marina Kounkel}
\affiliation{Physics \& Astronomy Department, Vanderbilt University, PMB 401807, 2301 Vanderbilt Place, Nashville, TN 37235}

\author[0000-0002-0458-7828]{Siyao Xu}
\altaffiliation{NASA Hubble Fellow}
\affiliation{Institute for Advanced Study, 1 Einstein Drive, Princeton, NJ 08540, USA}

\author[0000-0002-1253-2763]{Hui Li}
\altaffiliation{NASA Hubble Fellow}
\affil{Department of Astronomy, Columbia University, 550 West 120th Street, New York, NY 10027, USA}

\author[0000-0003-4158-5116]{Yong Zheng}
\affiliation{Department of Astronomy, University of California, Berkeley, CA 94720, USA}

\begin{abstract}

The interstellar medium (ISM) is turbulent on all scales and in all phases. In this paper, we study turbulence with different tracers in four nearby star-forming regions: Orion, Ophiuchus, Perseus, and Taurus. We combine the APOGEE-2 and Gaia surveys to obtain the full 6-dimensional measurements of positions and velocities of young stars in these regions. The velocity structure functions (VSFs) of the stars show a universal scaling of turbulence. We also obtain \halpha~gas kinematics in these four regions from the Wisconsin H-Alpha Mapper. The VSFs of the \halpha~are more diverse compared to the stars. In regions with recent supernova activities, they show characteristics of local energy injections and higher amplitudes compared to the VSFs of stars and of CO from the literature. Such difference in amplitude of the VSFs can be explained by the different energy and momentum transport from supernovae into different phases of the ISM, thus resulting in higher levels of turbulence in the warm ionized phase traced by \halpha. In regions without recent supernova activities, the VSFs of young stars, \halpha, and CO are generally consistent, indicating well-coupled turbulence between different phases. Within individual regions, the brighter parts of the \halpha~gas tend to have a higher level of turbulence than the low-emission parts. Our findings support a complex picture of the Milky Way ISM, where turbulence can be driven at different scales and inject energy unevenly into different phases.

\end{abstract}


\section{Introduction} \label{sec:intro}

The interstellar medium (ISM) is multi-phase and turbulent. Interstellar turbulence affects the velocity statistics on all scales from the diffuse ionized gas to the dense molecular gas. \citet{Larson81} compiled the internal velocity dispersion of nearby molecular clouds from the width of various emission lines, and found a power-law relation between the velocity dispersion and cloud size, with slope 0.38, between the 1/3 slope expected for turbulence dominated by solenoidal motions \citep{Kolmogorov41} and the 1/2 slope expected for shock-dominated turbulence \citep[e.g.,][]{2011PhRvL.107m4501G,Federrath13}.

Inside molecular clouds, massive O- and B-type stars create \ion{H}{2} regions by ionizing and heating the ISM with UV radiation \citep[see, e.g.,][]{1978ppim.book.....S}. 
Compression from supernova (SN) explosions can trigger further star formation \citep[][]{2012ARA&A..50..531K}. Interstellar processes such as SNe, stellar winds, ionizing radiation, and cosmic rays all play a role in maintaining the structure of the ISM, as well as in regulating star formation \citep[][]{Hennebelle12,2021MNRAS.504.1039R,2022MNRAS.tmp..939H,2022ApJ...927...94X}. Of these mechanisms, SN explosions are perhaps the most significant source of energy injection into the local ISM \citep[][]{2004RvMP...76..125M,Pad16}. 

Many numerical simulations have been carried out to study how SN explosions affect the surrounding ISM and star formation in the last decade. In general, they find that most of the kinetic energy from an SN explosion is transferred to the low-density region of the ISM in both single- \citep[see, e.g.,][]{2015ApJ...802...99K,Martizzi15,2015ApJ...814....4L} and multiple-SN events \citep[][]{Pad16}, suggesting that turbulence may not be evenly driven in different phases of the ISM.

Observational studies of ISM turbulence have been done with various gas tracers. 
Measurements of turbulence through density spectrum were carried out using the Wisconsin H-Alpha Mapper \citep[WHAM;][]{WHAM-North,WHAM-South}, which revealed a turbulent cascade consistent with Kolmogorov theory \citep[][]{2010ApJ...710..853C}. \citet[][]{2010ApJ...708.1204B} found that most of the \ion{H}{1} in the Small Magellanic Cloud (SMC) is subsonic or transonic, and concluded that turbulence in the SMC is dominated by hydrodynamical pressure over magnetic pressure. 

Statistical measurements of velocity fluctuations in the neutral ISM can be done using the Velocity Channel Analysis, the Velocity Coordinate Spectrum \citep{LP00,LP06}, the Delta-variance technique \citep{Stutzki1998,Ossenkopf02}, or the Principal Component Analysis \citep{Heyer04,Roman-Duval11} based on spectroscopic data. These analyses generally reveal power-law spectra of turbulent velocities \citep{Laz09rev,Chep10}.

Additionally, kinematics of Galactic \ion{H}{2} regions have been studied using the observations of ionized gas. Fabry-Perot interferometry was used to probe the \halpha~line in emission nebulae. \citet[][]{1970A&A.....8..486L} found a velocity-size relation that agrees with the 1/3 slope of Kolmogorov turbulence in M8, while \citet{1985ApJ...288..142R} and \citet[][]{1986ApJ...307..649J} found slopes of the structure functions closer to 1/2, indicating turbulence in the supersonic regime in both S142 and M17, respectively. 

Closer to our solar neighborhood, studies into the nature of turbulence in the Orion Nebula is an ongoing effort since the 1950s. \citet{1951ZA.....30...17V} observed a root-mean-square difference in emission line velocities that increases with larger angular separations. He reasoned that such relation could be explained by the Kolmogorov theory if the effective depth of the nebula was sufficiently small as to neglect projection effects. Subsequent studies, carried out by \citet{1958RvMP...30.1035M} and \citet{1988ApJS...67...93C}, disputes this conclusion. \citet[][]{1958RvMP...30.1035M} found an optical depth smaller than the value postulated by \citet[][]{1951ZA.....30...17V}, and attributed the failure of Kolmogorov theory to the compressibility of turbulence in the Orion Nebula, while \citet[][]{1988ApJS...67...93C} argued against a simple power-law relation, instead suggested energy injections at different scales. 

ISM turbulence has also been probed using ``point sources'' such as molecular cores. \citet{Qi12} developed statistical measurements of core-to-core velocity dispersion over a range of length scales , i.e., the core velocity dispersion technique. \citet{Qi18} find signatures for a turbulent power-law spectrum in the Taurus Molecular Cloud, and demonstrate that the statistics of turbulent velocities are imprinted in the velocities of dense cores \citep{Xu20}. 

One limitation of using gas tracers is they lack three-dimensional information on the position and the velocity information is also limited to only the line-of-sight.
In \citet[][hereafter Paper I]{Ha21}, we proposed a new method to measure turbulence in the ISM with young stars. Since stars are formed out of turbulent molecular clouds, we expected these young stars to inherit the turbulent kinematics from their natal clouds. Using six-dimensional position and velocity information of $> 1400$~stars in the Orion Complex, we found that the first-order velocity structure function (VSF) of young stars in loose groups exhibit characteristics of turbulence. Our method can therefore be used to probe the turbulent properties of molecular clouds in addition to analysis of gas kinematics and has been used in recent work in \citet{Taurus21,2022MNRAS.tmp..888Z}. 

In this work, we apply a similar analysis to a larger sample of molecular complexes, which include three of the nearest massive star-forming regions: Ophiuchus, Perseus, and Taurus. Additionally, we analyze the kinematics of \halpha~gas from the WHAM survey to study the connection between stars and the multi-phase ISM.

In Section \ref{sec:data_method}, we describe the data used in this analysis, the reduction pipeline for the \halpha~kinematic survey, and how we compute the VSFs using stars and \halpha. In Section \ref{sec:results}, we present the VSF of young stars and multi-phase gas of all four regions. 
In Section \ref{sec:discussion}, we discuss sources of uncertainty associated with our analysis and the nature of turbulence in the Milky Way ISM. We summarize out results in Section \ref{sec:conclusions}.

\begin{figure*}
    \centering
    \includegraphics[width=0.37\linewidth,trim=0cm 0cm 0cm 0.2cm, clip=true]{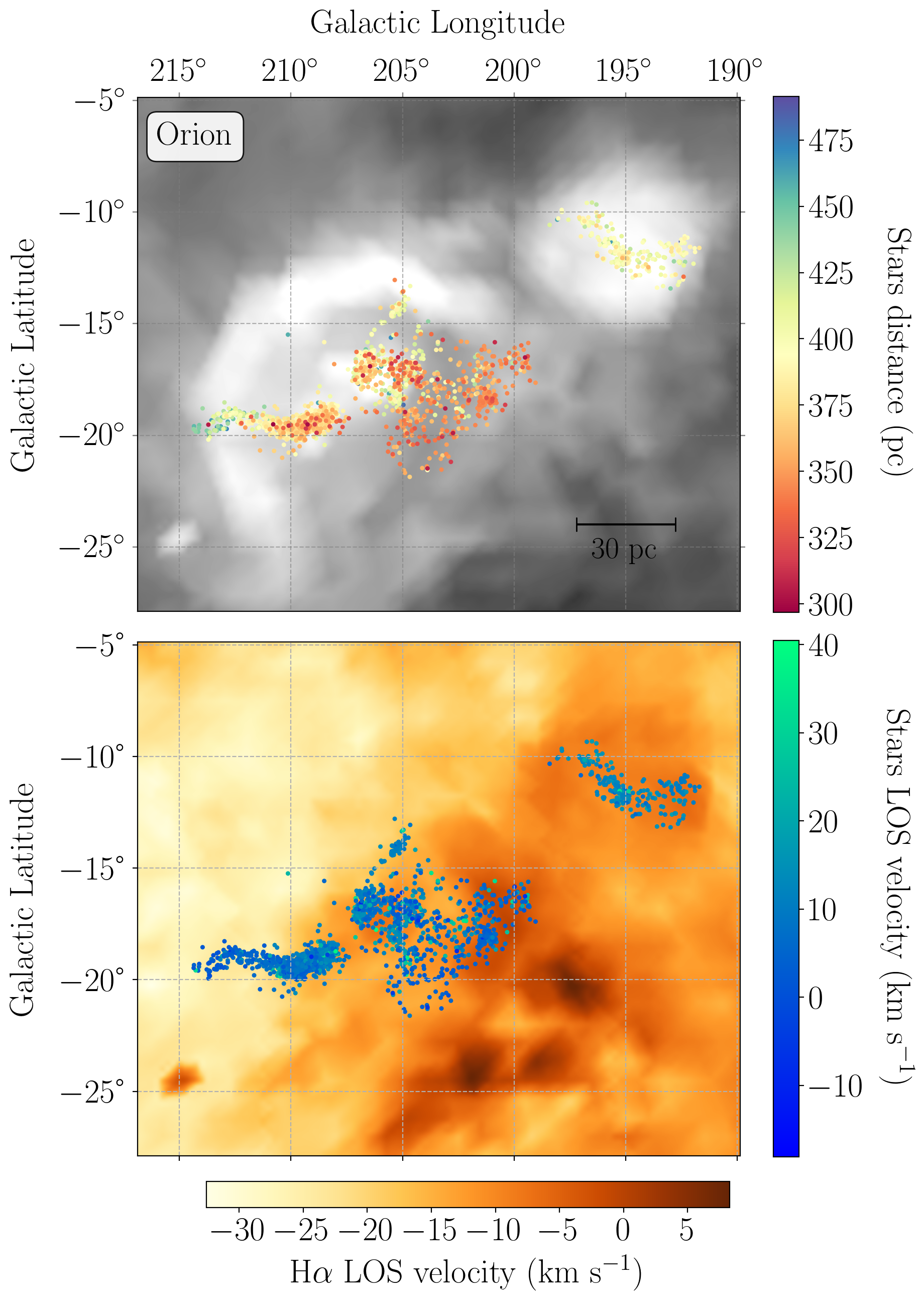}
    \includegraphics[width=0.312\linewidth,trim=1.5cm 0cm 0cm 0.1cm, clip=true]{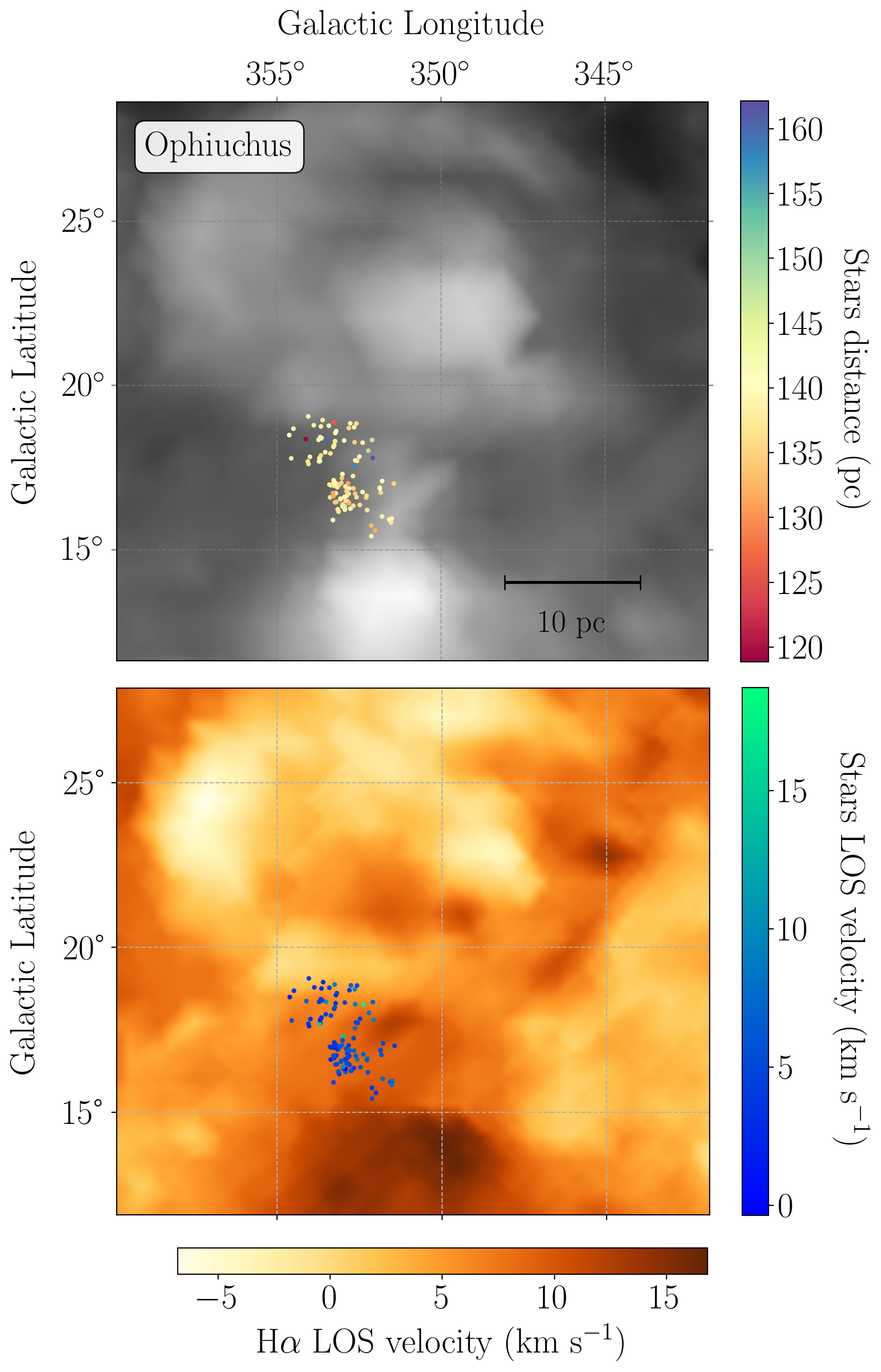}
    \includegraphics[width=0.302\linewidth,trim=1.6cm 0cm 0.1cm 0.1cm, clip=true]{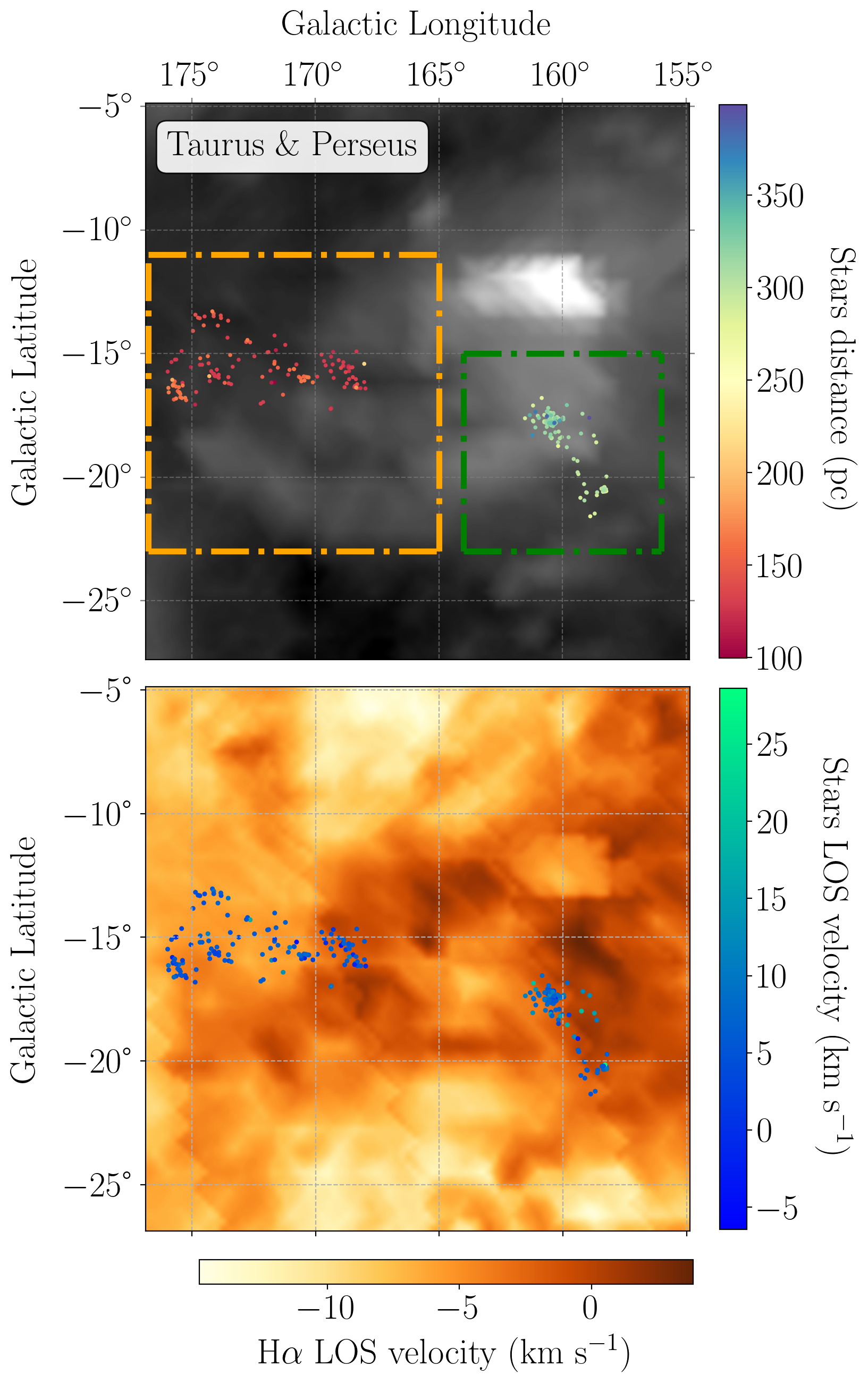}
    \caption{Top panels: positions of young stars color-coded by their distances in Orion, Ophiuchus, Taurus-Perseus (from left to right). Background shows \halpha~flux from the WHAM survey (lighter is brighter). 
    The orange and green boxes denote the areas where \halpha~is analyzed for Taurus and Perseus, respectively.
    Bottom panels show the line-of-sight velocities of the young stars (in the LSR frame) in these regions. The background shows the \halpha~line-of-sight centroid velocities.}
    \label{fig:allsky}
\end{figure*}
\begin{deluxetable*}{ccccccccc}
\tablecolumns{9}
\tablecaption{Properties of Star-Forming Regions}
\tablehead{
\colhead{Region} &
\colhead{Number} &
\colhead{Number of} &
\colhead{Average} &
\colhead{GLON} &
\colhead{GLAT} &
\colhead{Distance} &
\colhead{$v_{\rm LSR}$~range} &
\colhead{$v_{\rm LSR,~err}$~cut}\cr
\colhead{} &
\colhead{of stars} &
\colhead{stars used} &
\colhead{distance (pc)} &
\colhead{(\textdegree)} &
\colhead{(\textdegree)} &
\colhead{range (pc)} &
\colhead{($\mathrm{km~s}^{-1}$)} &
\colhead{($\mathrm{km~s}^{-1}$)}
}
\startdata
Orion & 2647 & 2468 & 410 & $190,217$ & $-28,-5$ & $300,500$ & $-10,40$ & $10$\\
Ophiuchus & 113 & 107 & 140 & $342,360$ & $12,28$ & $120,170$ & $-30,30$ & $2$\\
Perseus & 159 & 138 & 320 & $156,164$ & $-23,-15$ & $250,400$ & $-6,30$ & $10$\\
Taurus & 157 & 139 & 140 &  $165,177$ & $-23,-11$ & $97,200$ & $-5,25$ & $5$
\enddata
\tablecomments{Number of stars are all the stars with 6D information. Number of stars used are the stars in our analysis after excluding outliers. GLON (Galactic Longitude) and GLAT (Galactic Latitude) ranges denote the borders of windows used to compute the \halpha~VSF. The distance ranges, $v_{\rm LSR}$~ranges, and $v_{\rm LSR,~err}$~cuts are applied to the stars to exclude outliers.}
\label{tab:properties}
\end{deluxetable*}

\vspace*{-0.3in}
\section{Observations, Data reduction, and Methods} \label{sec:data_method}
\subsection{Observations and Data Analysis} \label{sec:alldata}
\subsubsection{Stellar observations} \label{subsec:stars_data}

We obtain the locations, parallax distances, and proper motions of stars from the Gaia mission \citep{GaiaMission} in four nearby star-forming regions: the Orion Molecular Cloud complex (Orion), the \rophi~dark cloud (Ophiuchus), the Perseus molecular cloud traced by two clusters, NGC 1333 and IC 348 (Perseus), and the Taurus molecular clouds (Taurus). The Gaia catalog includes five-parameter astrometric measurements (plane-of-sky coordinates, proper motions, and parallax distances) of over 1.46 billion stars in the Milky Way. Only a very limited subset ($\sim 7$ million) has line-of-sight velocities measured by Gaia. The five-dimensional astrometric data for our four regions are obtained from Gaia's third early data release \citep[EDR3;][]{GaiaEDR3}. 

The line-of-sight velocities of the stars in our study were observed in the near infrared with the Apache Point Observatory Galactic Evolution Experiment (APOGEE) spectrograph, mounted on the 2.5 m Sloan Foundation Telescope of the Sloan Digital Sky Survey. APOGEE-2 Data Release 17 (DR17) collected high resolution spectral image of over 657,000 stars in the Milky Way \citep{Gunn06,Blanton17,2022ApJS..259...35A}. Combining these two surveys provides six-dimensional (6D) information (3D positions and 3D velocities) of stars for the four molecular clouds.

The stars in our study were assigned to each of the four region by \citet{KounkelMW1}. They performed a clustering analysis on Gaia DR2 sources in the Milky Way using Python implementation of HDBSCAN \citep[Hierarchical Density-Based Spatial Clustering of Applications with Noise;][]{2017JOSS....2..205M}. We crossmatch this catalog with Gaia EDR3 and APOGEE-2 to obtain 6D astrometries of the stars. 

Then, we exclude a small number of stars whose parallax distances $d$, line-of-sight velocities $v_{\rm LSR}$, and line-of-sight velocity errors $v_{\rm LSR,~err}$~are more than 4$\sigma$~deviated from the mean of the distribution. In Taurus, one star with very high reported proper motion ($v_{\rm LSR,~RA} > 100\mathrm{~mas~yr}^{-1}$) is excluded.
The ranges of $d$~and $v_{\rm LSR}$, and the threshold of $v_{\rm LSR,~err}$~that we use, as well as the total number of stars before and after the cut are specified in Table \ref{tab:properties}. We note, however, that the exact cut of these ranges do not affect the conclusion of this work.

\subsubsection{\halpha~observations} \label{subsec:Ha_data}

The \halpha~data in our study comes from the WHAM survey \citep[][]{WHAM-North,WHAM-South}. We obtained the data cubes from the ``DR1-v161116-170912" release, which comprises more than 49000 spectra divided into an irregular grid with a 1\textdegree~seeing resolution and a 0.25\textdegree~sampling resolution. The intensity of the \halpha~emission is recorded in a 3-dimensional array encoding its Galactic longitude, Galactic latitude, and line-of-sight velocity ($-170$ km s$^{-1} < v_{\rm LSR} < 160$ km s$^{-1}$). We fit a Gaussian distribution to the flux profile along individual lines of sight. Then we take the velocity at the peak of the distribution as the centroid velocity and the uncertainty of the fit as the uncertainty. We also correct for Galactic rotation following the procedure in \citet{Qu20}.

Figure \ref{fig:allsky} shows the \halpha~map of the four regions in our analysis, with stars overplotted as foreground dots. The top panels show the integrated line-of-sight fluxes in log scale. In the foreground, each star is color-coded by their distance from Earth. We note that although Perseus and Taurus are very close to each other on the sky, they are physically separated by $\sim 150$~pc along the line of sight. In the bottom panels, we plot the centroid velocity of the \halpha~in the background, and the line-of-sight velocities of stars in the foreground.

\subsection{Data analysis} \label{sec:analysis}

We compute the first-order VSF of stars in each region. The VSF is related to the kinetic energy power spectrum of a velocity field, and is expressed as a two-point correlation function between the absolute velocity differences $|\delta v|$ versus the physical separation $\ell$. For a turbulent velocity field, we expect the VSF to have a characteristic power-law scaling with slope 1/3 for incompressible turbulence dominated by solenoidal motions \citep[][]{Kolmogorov41}, or a slope 1/2 for supersonic turbulence dominated by compressive motions. For the Milky Way molecular clouds, \citet{Larson81} found a relationship between cloud size and velocity dispersion to be $\sigma = 1.1~R^{0.38}$.

The 3D VSF of stars is computed as follows. First, the positions, distances, radial velocities, and proper motions of stars are translated into galactocentric Cartesian coordinates and velocities. Then, for each pair of stars, the absolute velocity vector difference $|\delta \vec v|$ and the distance $d$ between them is calculated. For individual clouds, these distances are sorted into logarithmic bins of $\ell$, and $|\delta \vec v|$ values within a bin are averaged out to obtain \vsf. To estimate uncertainties, we first generate 1000 random samples of stellar parameters following a Gaussian distribution around the reported measurements and with a 1$\sigma$ scatter equal to the reported uncertainties. Then at each $\ell$, we take the 1$\sigma$ value of the 1000 \vsf~to be the uncertainty.

For the WHAM \halpha~data, we first define a rectangular box containing the stars in each of the four regions as specified in Table \ref{tab:properties} and shown in Figure~\ref{fig:allsky}. Our results are not sensitive to the exact size of the box (see Appendix \ref{sec:windowsize} for detailed discussions). We compute the first-order VSF of each region using the projected position-position-velocity information. The resultant VSF is $\langle |\delta v| \rangle_{\rm proj}$~versus angular separations $\theta$. Then, we assume a constant distance equal to the mean distance of stars in each region (see Table~\ref{tab:properties}) and translate $\theta$~(\textdegree) to $\ell_{\rm proj}$~(pc). Since three degrees of freedom are missing in this projected VSF calculation (the proper motions of gas and its exact distance), we scale $\langle |\delta v| \rangle_{\rm proj}$~and $\ell_{\rm proj}$ by a constant factor, assuming an isotropic turbulent velocity field. 

\begin{equation}
    \langle|\delta \vec v|\rangle = \sqrt{3}~ \langle|\delta v|\rangle_{\rm proj},
\end{equation}
\begin{equation}
    \ell = \sqrt{\frac{3}{2}}~\ell_{\rm proj}.
\end{equation}

\section{Results} \label{sec:results}

\begin{figure}
	\centering
	\includegraphics[width=\linewidth]{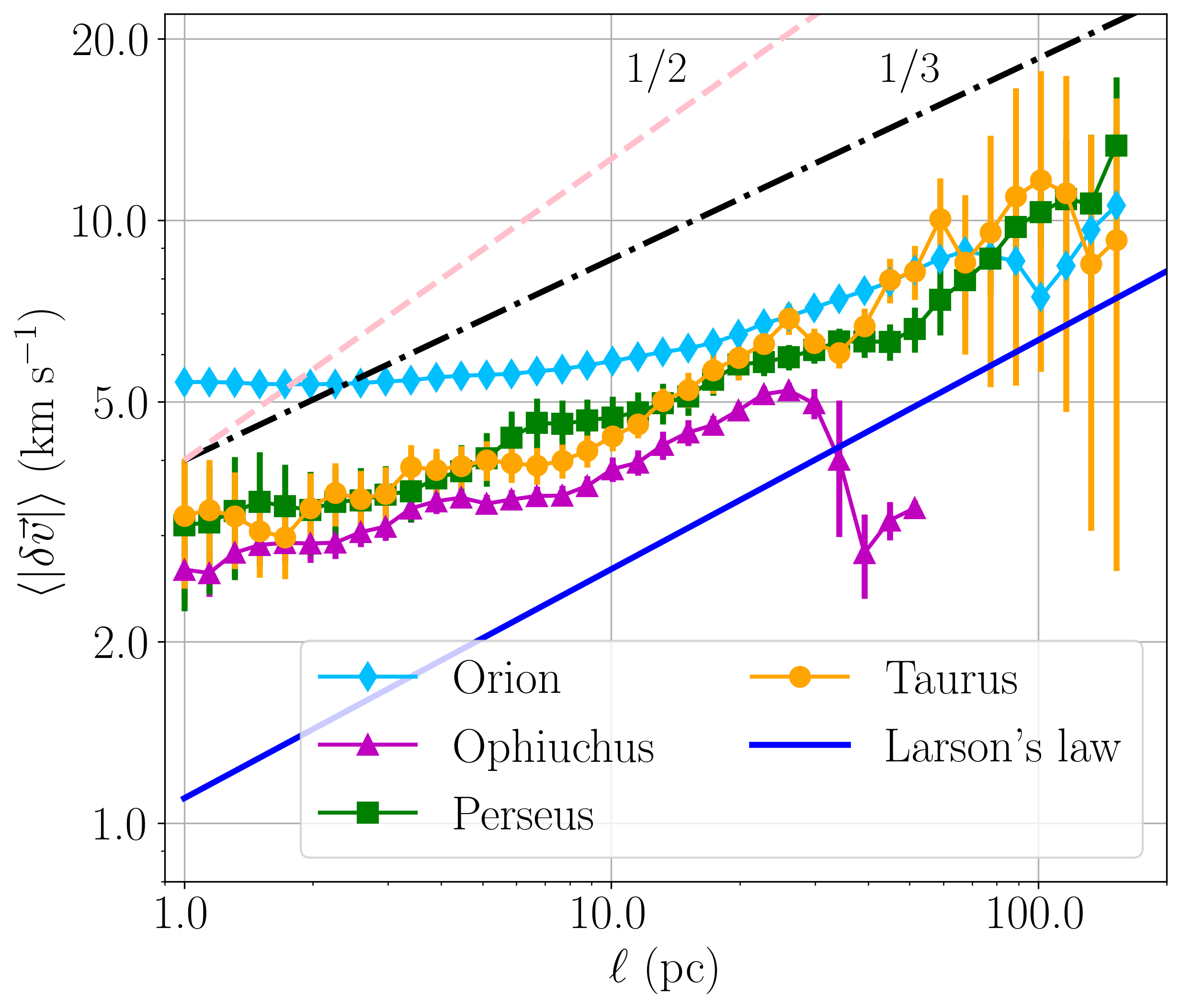}
	\caption{The first-order VSFs of the stars in Orion (light blue), Ophiuchus (magenta), Perseus (green), and Taurus (orange). For reference, we plot a solid blue line for Larson's relation, a black dotted-dashed line for Kolmogorov turbulence, and a pink dashed line for supersonic turbulence. The vertical lines represent the uncertainties in \vsf~ at each $\ell$.}
	\label{fig:starsvsf}
\end{figure}

\subsection{The Universality of ISM Turbulence Traced by Young Stars}\label{sec:starvsf}
We first examine the velocity statistics of the four star-forming regions: Orion, Ophiuchus, Perseus, and Taurus. Figure~\ref{fig:starsvsf} shows the VSFs traced by stars of these four regions. For reference, we also plot the Larson's law of velocity dispersion vs. cloud size, the 1/3 slope of Kolmogorov turbulence, and the 1/2 slope of supersonic turbulence. Immediately, we recognize a remarkable universality among the VSFs of different regions. Over a large dynamical range, the magnitude and slopes of the VSFs show a general agreement with Larson's law (see Table~\ref{tab:slopes} for the best-fit slopes of the VSFs). This suggests that the motion of young stars in the Milky Way show characteristics of turbulence, and supports our previous findings in Paper I that turbulence in the ISM can be traced with young stars.

At $\ell \lesssim 10$~pc, the slopes of the VSF of all four regions are flatter than Larson's law, as well as Kolmogorov (slope 1/3) turbulence. We discuss the possible cause of this flattening in Section~\ref{sec:star_discussion}.

\subsection{The VSFs of Individual Regions} \label{subset:eachregion_result}

\begin{figure*}[ht!]
    \centering
    \includegraphics[width=0.48\linewidth]{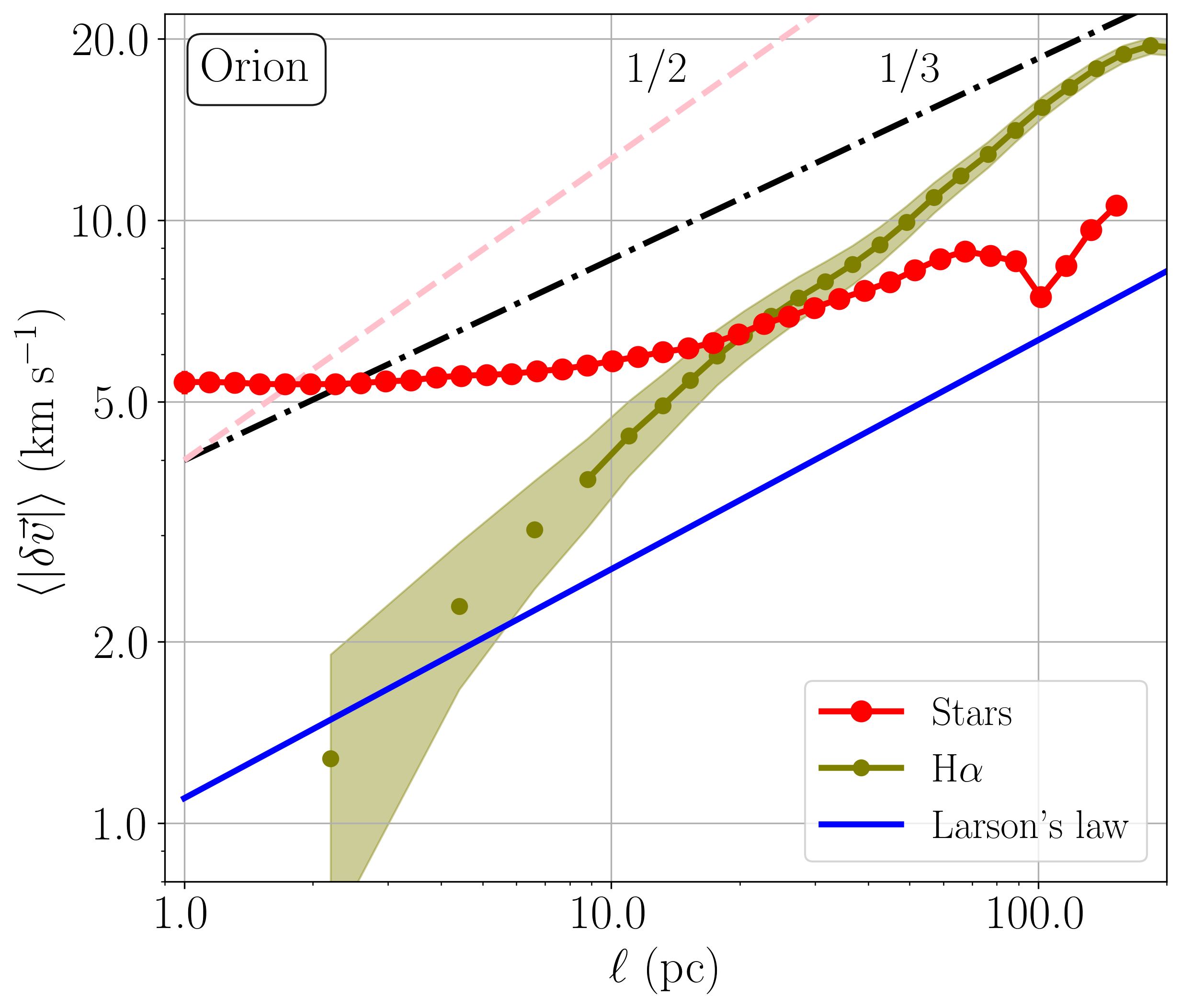}
    \includegraphics[width=0.48\linewidth]{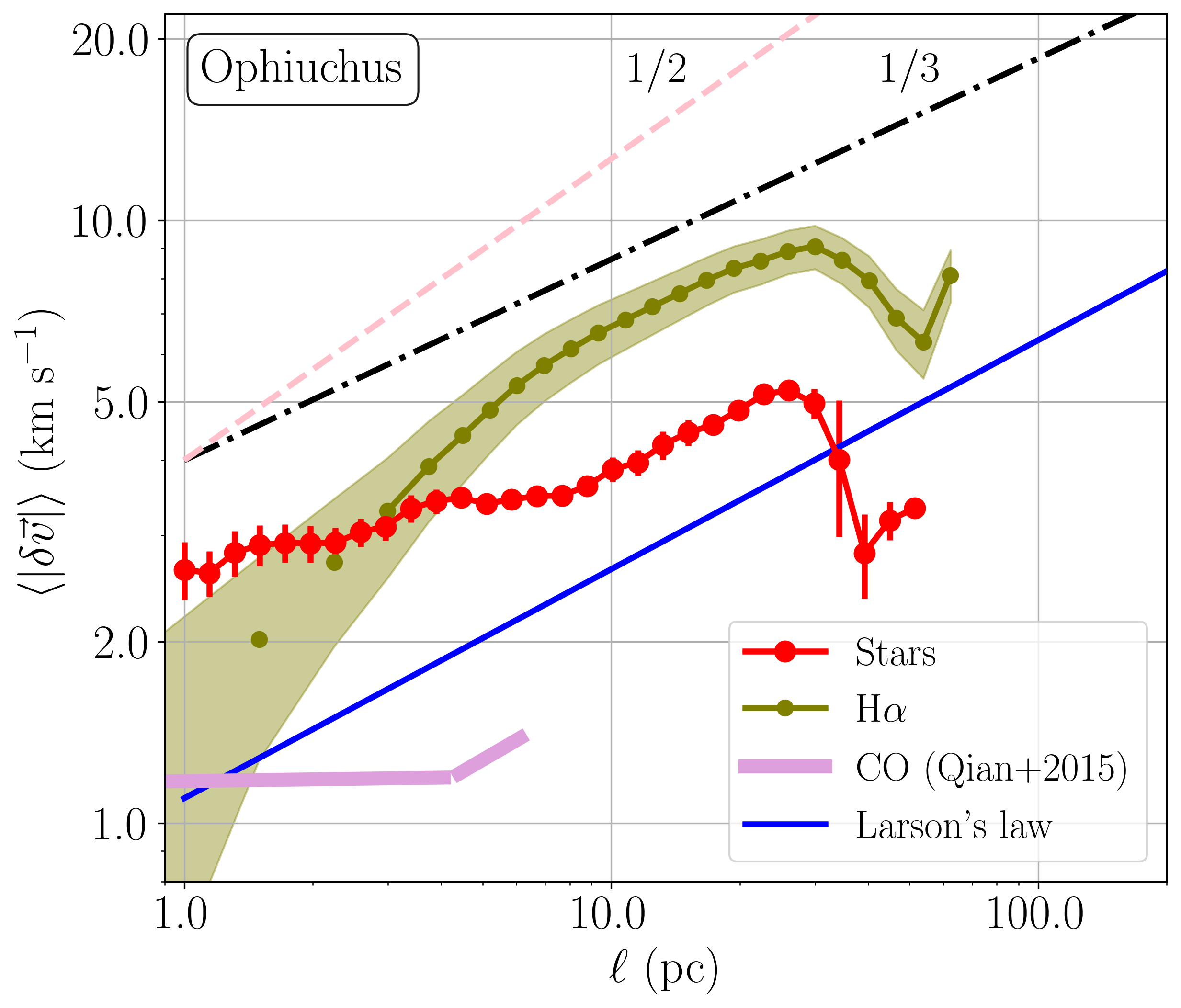}
    \includegraphics[width=0.48\linewidth]{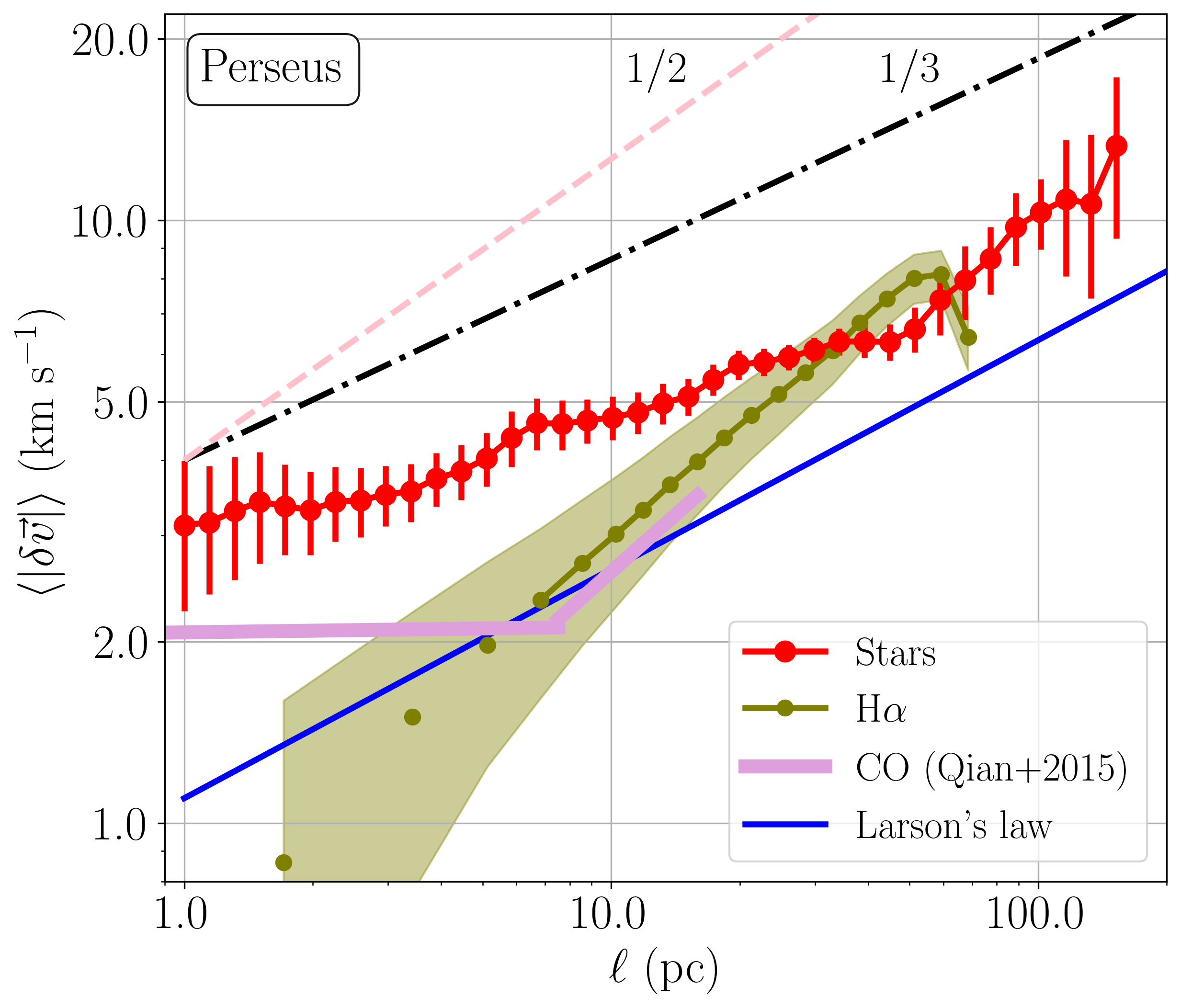}
    \includegraphics[width=0.48\linewidth]{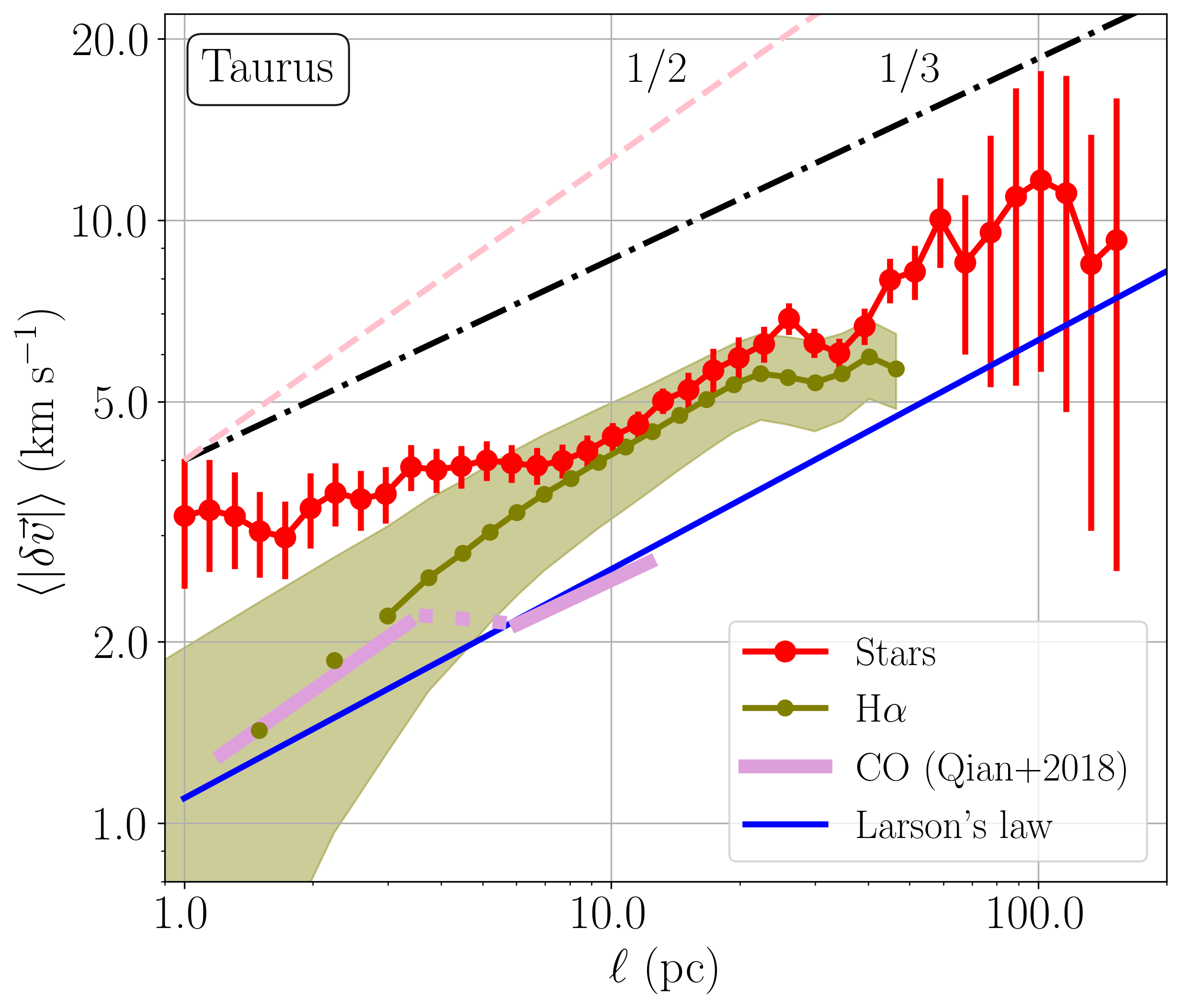}
    \caption{The first-order 3D VSF of stars (red) and H$\alpha$ (olive) for Orion, Ophiuchus, Perseus, and Taurus. The shaded region around the \halpha~VSF is its associated uncertainty. At small $\ell$, the dots showing the VSF traced by \halpha~are not connected to indicate the seeing limit of the WHAM survey ($\theta < 1$\textdegree). The slopes of the VSFs of stars and of \halpha~are specified in Table \ref{tab:slopes}. The VSF of \halpha~in Ophiuchus~can also be fitted with a broken power-law, with slope $0.58 \pm 0.24$ for $3~\mathrm{pc} < l < 9$~pc and slope $0.29 \pm 0.08$ for $9~\mathrm{pc} < l < 30$~pc. For comparison, we also include CO (violet) analysis from \citet{Qi15,Qi18} for three of these regions.}
    \label{fig:stars_gas}
\end{figure*}

Figure \ref{fig:stars_gas} shows the VSFs of young stars as well as gas in each of the four regions: Orion, Ophiuchus, Perseus, and Taurus. The best-fit slopes of these VSFs are lists in Table~\ref{tab:slopes}.

\subsubsection{Orion}

The first-order VSF of Orion traced by stars has been studied in detail in Paper I. We only reiterate some of the main points here and address the differences caused by the addition of $\sim 1000$~stars compared with Paper I. 
Throughout the inertial range, the magnitude of $\langle |\delta \vec{v}| \rangle$~is higher than Larson's law. At small scales ($1~\mathrm{pc} \lesssim \ell \lesssim 20$~pc), the number statistics is dominated by stars in the Orion Nebula Cluster (ONC), which flattens the VSF (see Section~\ref{sec:star_discussion} for more discussion on the small-scale flattening). 
At ranges from $20$~pc to $\sim 70$~pc, Orion D, which is a loose stellar association at the front of the Orion Complex along our line of sight, dominates the statistics, and it exhibits a power-law scaling, which indicates an imprint of turbulence from their natal molecular clouds. The VSF of Orion without considering the ONC would be steeper at small scales, which renders it more in-line with the VSF of other regions (see Figure \ref{fig:starsvsf}).
At $\sim 70$~pc, there is a bump in the structure function, and we proposed that this is an evidence supporting the existence of a SN explosion at the heart of Orion 6 Myr ago \citep{Kounkel20}.
Paper I used Gaia DR2 only. The newly added stars in this study from Gaia EDR3 have larger overall uncertainty in their parametric solutions. Therefore, the noise floor for Orion at small scales is boosted from $\sim 3$~km s$^{-1}$ to $\sim 5$~km s$^{-1}$. This slightly elevates the overall VSF compared to Paper I. 

\halpha~emission in Orion mainly comes from the Barnard's loop and the $\lambda$~Ori bubble (see Figure~\ref{fig:allsky}), which are thought to be created by SN explosions \citep{Kounkel20}. We see the effect of such energy injection in the VSF traced by \halpha. At $\ell \gtrsim 20$~pc, the amplitude of the \halpha~VSF is higher than the VSF of young stars. Over the whole dynamical range we probe here, the \halpha~VSF has a roughly constant slope of 1/2. 

The VSF traced by \halpha~shows a bump at $\ell \sim 150$~pc. This is indicative of a local energy injection, similar to those seen in the VSF traced by young stars. However, we caution that this scale is very close to the size of the \halpha~box, and the bump can also be an artifact \citep[see, e.g.,][]{2022MNRAS.510.2327M}. We discuss this effect in more details in Appendix \ref{sec:windowsize}. 
At very small scales ($\theta = 1^{\rm o}$, corresponding to $\ell \sim 7$~pc), our analysis is limited by the resolution of the WHAM survey \citep[][]{WHAM-North}. We discuss the seeing effect and other uncertainties related to \halpha~VSF in Section~\ref{sec:halpha_discussion}.

\subsubsection{Ophiuchus}

The young stars in Ophiuchus~has a VSF similar to that of the Orion Complex. At small $\ell$, the structure function is flatter than 1/3, and then steepens to $\sim$ 1/3 beyond 10 pc up to $\sim 25$~pc. The \halpha~VSF in Ophiuchus has slope close to 1/3 and a higher amplitude than the VSF of the stars. Both VSFs show clear bumps at $\sim 25$~pc, which we interpret as evidence of local energy injection.

In Orion, such an energy injection is associated with a past SN explosion as we discuss in detail in Paper I. If the energy injection in Ophiuchus is also caused by a SN, we estimate that the event took place around $25\,\mathrm{pc} / v_{\rm region} \sim 2.5$~Myr ago, where we use the characteristic \halpha~velocity $v_{\rm region}\sim 10$~km s$^{-1}$. Such an estimate is broadly consistent with observations of runaway stars which trace back to a SN explosion in the region $\sim 2$~Myr ago \citep[][]{Ophi_SN20}. 

We note that the scale of the bump is close to the size of the region containing the young stars in our analysis. Therefore, it is technically possible that the bump in the VSF of young stars is only an artifact caused by the limited sample size. In Appendix \ref{sec:Upper_Sco}, we analyze the entire Upper Sco region. We find that as long as we apply our analysis to young stars ($< 2$~Myr), the results are consistent, while the VSFs of older stars do not show this feature. In addition, the SN scenario is also supported by the bump in the \halpha~VSF.

The cold and dense molecular cores, where stars are born out of, emits CO. \citet{Qi15, Qi18} applied the core-velocity-dispersion \citep[CVD;][]{Qi12}~method to analyze turbulence in star-forming regions. We translate their best-fit CVD to a VSF of CO by $\mathrm{VSF} = \sqrt{\frac{2}{\pi}}~\mathrm{CVD}$. 

In Ophiuchus, the amplitude of CO VSF (pink solid lines in Figure \ref{fig:stars_gas}) is much lower than the VSF of young stars and \halpha. It is possible that the recent energy injection has affected the warm phase of the ISM more than the cold phase (see discussion in Section~\ref{sec:nature_ISM}). It is also possible that the low amplitude is due to spacial variation of ISM turbulence (see discussion in Appendix \ref{sec:windowsize}). \citet[][]{Qi15} only examined the smaller scale velocity structures ($\ell \lesssim 10$~pc), where there is no reliable component that we can compare against (due to relaxation of stars and resolution limit of \halpha). 

\subsubsection{Perseus}
At length scales below 10 pc, the VSF of young stars in Perseus is very similar to that of Ophiuchus~(see Figure \ref{fig:starsvsf}). Since we include two clusters in our analysis of Perseus: NGC 1333 and IC 348, the VSF is flattened from internal dynamical relaxation. The existence of subgroups and sub-clusters are likely the cause of the small wiggles in the VSF. At larger scales, the VSF is generally consistent with Larson's law.

The VSF traced by \halpha~here has an amplitude similar to the VSF traced by young stars but a steeper slope ($\sim$ 1/2). VSF traced by CO matches the \halpha~ VSF at $\ell > 6$~pc, but flattens on smaller scales due to projection effects for thick clouds \citep[][]{Qi15, Xu20}.

\subsubsection{Taurus}

Taurus is overall very similar to Perseus. The VSFs of young stars, \halpha~and CO generally agree with each other, as well as Larson's Law.

There is no prominent bump detected in the VSF in both Perseus and Taurus, which is consistent with the findings of \citet[][]{Taurus21}~that there has been no SN activities in the last 5 Myr in these regions. However, there are wiggles scattered throughout the length scales, likely because Taurus contains several small dense groups that are primarily homogeneously distributed throughout the region.

\begin{deluxetable}{ccccc}
\tablecolumns{5}
\tablecaption{Slopes of the VSFs}
\tablehead{
\colhead{Region} &
\colhead{Slope of} &
\colhead{Fitting} &
\colhead{Slope of} &
\colhead{Fitting}\cr
\colhead{} &
\colhead{stars VSF} &
\colhead{range (pc)} &
\colhead{\halpha~VSF} &
\colhead{range (pc)}
}
\startdata
Orion & $0.25 \pm 0.01$ & $12 - 75$ & $0.55 \pm 0.02$ & $9 - 150$\\
Ophiuchus & $0.34 \pm 0.03$ & $7 - 28$ & $0.36 \pm 0.03$ & $3 - 30$\\
Perseus & $0.22 \pm 0.03$ & $5 - 80$ & $0.61 \pm 0.08$ & $7 - 55$\\
Taurus & $0.34 \pm 0.03$ & $7 - 70$ & $0.42 \pm 0.12$ & $3 - 25$
\enddata
\tablecomments{Inside the fitting ranges, the VSFs are fitted with an equation: \vsf~$= A \cdot \ell^{B}$, where $B$~is the slope. The uncertainties reported with each slope is obtained from the covariance matrix, taking into account the uncertainties in \vsf~shown in Figure~\ref{fig:stars_gas}. We choose the fitting ranges such that the uncertainties are low and there is a consistent trend in the slope of the VSF.}
\label{tab:slopes}
\end{deluxetable}

\section{Discussion} \label{sec:discussion}

\subsection{Uncertainties and biases in turbulence traced by young stars} \label{sec:star_discussion}

We discussed various biases and uncertainties associated with measuring turbulence from stars in Paper I, including dynamical relaxation, binary contamination, and drifting. In this section, we reiterate the main points and discuss additional mechanisms that can affect the shape of the VSF.

Stellar interactions can cause dynamical relaxation, which erases the memory of turbulence and flattens the VSF. Similar to Orion's loose groups discussed in Paper I, the relaxation times of Ophiuchus, Perseus, and Taurus are estimated to be tens of Myrs, much longer than their age \citep[for a detailed discussion of the age of each region, see][]{2008hsf2.book..351W,Taurus21,2021MNRAS.503.3232P}. Thus dynamical relaxation has not affected the overall shape of the VSF for the three regions. However, there are some small, compact star clusters in these regions, which can contribute to the flattening at small scales ($\ell \lesssim 10$~pc, see Figure \ref{fig:starsvsf}).

Another source of uncertainty originates from the drifting of stars along their natural trajectories over time. Stars born with high velocity differences drift apart at faster rates than stars born with low velocity differences, which steepens the VSF roughly over the the crossing time ($t_{\rm cross}$) of the cloud. Similar to the star groups in Orion that we discussed in Paper I, $t_{cross}$ of Ophiuchus, Perseus, and Taurus are all larger than their current age, and thus this effect is unlikely to have significantly altered the shape of the structure functions.

Low-velocity binary systems and contamination from field stars can act like noise in our analysis, which also tend to flatten the VSF as is discussed in Paper I.

Additionally, other stellar feedback processes such as winds, jets, and radiation can also inject energy on small scales \citep[e.g.,][]{2009ApJ...694L..26G, Offner15, Hui2019, Rosen21, 2022MNRAS.512..216G}. This can contribute to the flattening of the VSFs at small scales, provided that these stars are massive ($> 8~M_{\odot}$) and are within $\sim 1$~pc of each other \citep[][]{Rosen21}. 

\citet{Runaway_Stutz} quantified an ejection mechanism through which protostars decouple from the molecular gas. They found that the undulation of star-forming filaments in Orion A ``slingshots" protostars out of the filament at $\sim 2.5$~km~s$^{-1}$, which is also at the level where the VSF flattens in Ophiuchus, Perseus, and Taurus. 

All sources of uncertainties listed above can contribute to the flattening of the VSF at small scales ($\ell \lesssim 10$~pc). Therefore, we only consider young stars to be reliable tracers of ISM turbulence above $\sim 10$~pc.

\subsection{ISM Turbulence traced by \halpha}
\label{sec:halpha_discussion}

In Figure \ref{fig:starsvsf} and Figure \ref{fig:stars_gas}, we find that while the VSFs of stars are broadly consistent across the regions, the VSFs of \halpha~are more diverse. 

In regions with recent supernova explosions (Orion and Ophiuchus), the VSF of \halpha~gas has a higher amplitude than that of stars. Whereas in Perseus and Taurus, the \halpha~flux is lower (Figure~\ref{fig:allsky}) and the \halpha~VSF is more in line with the VSFs of stars and CO.

\begin{figure}
    \centering
    \includegraphics[width=\linewidth]{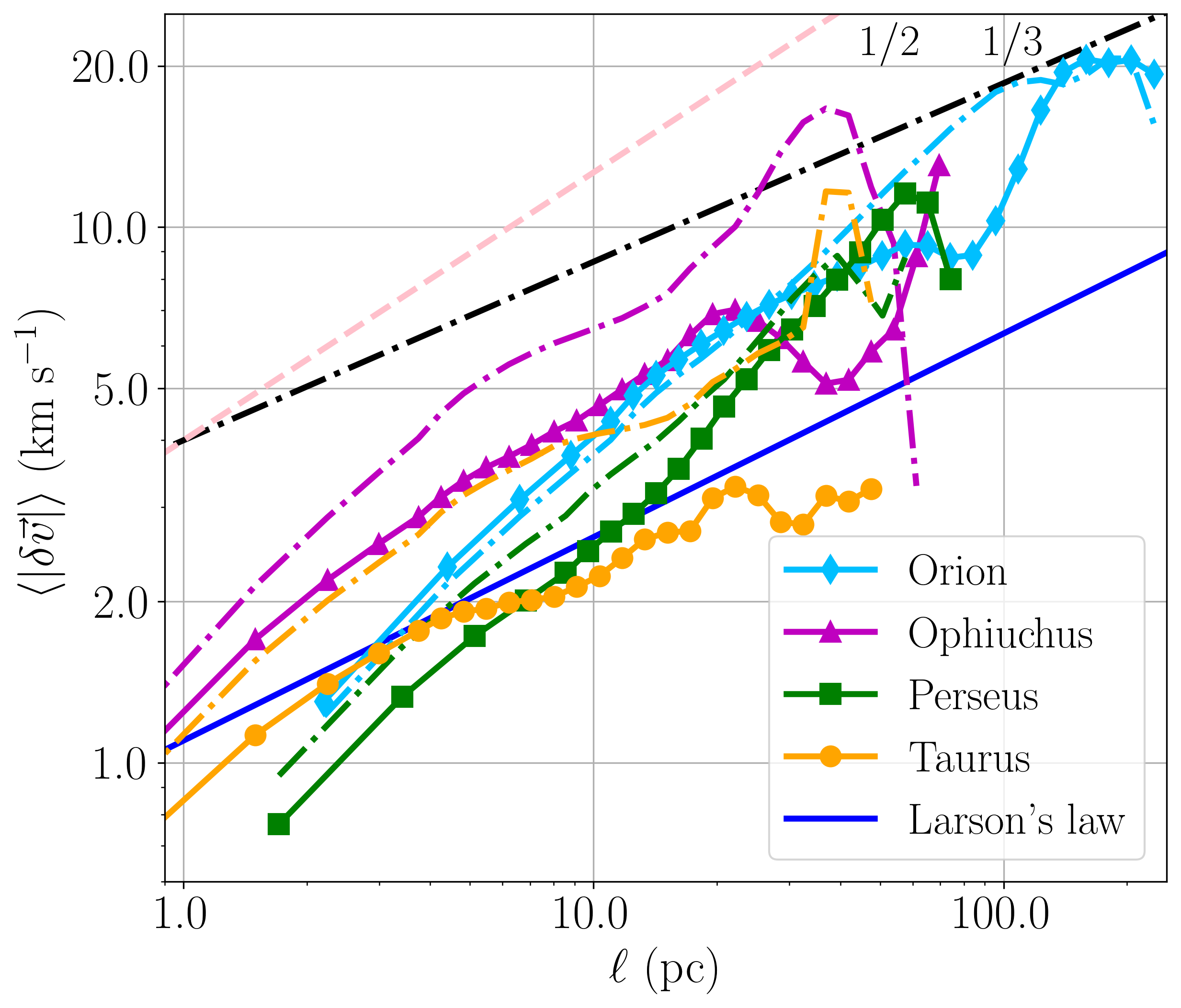}
    \caption{VSF of four regions traced by \halpha. Solid lines denote the VSF computed for pixels with low emission, and dotted-dashed lines denote the VSF computed for pixels with high emission.}
    \label{fig:emission}
\end{figure}

Within each individual region, we divide the pixels in half based on the \halpha~brightness--
the high-emission half with all the pixels brighter than the median and the low-emission half with all the pixels dimmer than the median. Figure \ref{fig:emission}~shows the VSFs of high-emission \halpha~ (dotted-dashed line) and low-emission \halpha~(solid line) within each region. In general, the VSF of high-emission \halpha~has a higher amplitude compared to that of the low-emission \halpha. 

One possible explanation for this is that bright \halpha~comes from photo-ionization by young massive stars \citep{2022MNRAS.tmp..796T}. Therefore, it traces the most intense stellar feedback regions and shows a higher VSF amplitude. Whereas the low-emission \halpha~ comes from primarily collisional ionization. Note that Orion has its high- and low-emission \halpha~ VSFs essentially overlapping for the most part. This is likely because star formation in Orion is the most intense among all the regions studied here \citep[see Table 2 of][]{2010ApJ...724..687L}, and thus most of the \halpha~is photo-ionized by massive stars.

There are many uncertainties and potential biases associated with the \halpha~analysis. One uncertainty comes from atmospheric seeing. Although the WHAM data cubes provide us with \halpha~line-of-sight measurements every 0.25\textdegree, the seeing limit of the survey is reported as 1\textdegree. At separations $\theta \leq 1$\textdegree~(or $\ell_{\rm proj} \leq d \cdot 1$\textdegree), we treat the VSF traced by \halpha~as unresolved (represented by the unconnected dots in Figure \ref{fig:stars_gas}). We note that seeing might not significantly alter the overall VSF, as suggested by \citet{Li20}. Future studies with mock simulated \halpha~observations may help properly evaluate the effects of seeing. 

There are also uncertainties with the distance to the \halpha~gas. The bright \halpha~emissions in Orion and Ophiuchus~are likely associated with the star forming regions. However, since Perseus and Taurus do not emit intense \halpha~ (see top right panel of Figure \ref{fig:allsky}), it is uncertain whether the centroid velocities found in these two regions are at the exact same distance as the stars. 

Another bias can come from the projection effects. When analysing the VSF of \halpha, we assume that all the gas is located at the same distance, effectively treating it as a 2D structure. This can flatten the VSF for thick clouds, as is discussed in detail in \citet{Qi15, Xu20, 2022MNRAS.510.2327M}. All the \halpha~VSFs in our analysis is either as steep as 1/3 or steeper. This suggests that either most of the \halpha~comes from a thin sheet of gas, or the true VSF has a even steeper slope.

Another potential bias related to projection is that we are trying to infer 3D turbulence using projected 2D information. This can introduce uncertainties as the ISM is magnetized. In the presence of magnetic fields, turbulence is no longer isotropic \citep{2015MNRAS.452.2410S, 2021ApJ...915...67H}.

We find that our results are not sensitive to the rotation model (see detailed discussions in Appendix \ref{sec:rotation}). 

\subsection{The Complex Nature of ISM Turbulence}
\label{sec:nature_ISM}

The ISM is multi-phase in nature. The kinematic coupling between these phases is not well understood \citep[for a review of the multi-phase ISM model, see][]{2005ARA&A..43..337C}. More specifically, it is suggested that each phase possesses varying levels of turbulence, where ionized gas has been observed to have higher velocity dispersion than molecular gas in both high- and low-redshift galaxies \citep[][]{2021ApJ...909...12G,2021arXiv211109322E}. In our study, we observe a similar trend comparing the VSF of \halpha~and other tracers of turbulence.

In Orion and Ophiuchus, there is a large deviation between the level of turbulence seen by different tracers. The VSFs traced by \halpha~show the largest amplitude at large scales, followed by the VSFs traced by stars, and finally by CO. One possible explanation is the difference in the transport of energy and momentum from SNe into different phases in these regions, as mentioned in Section \ref{subset:eachregion_result}. In an inhomogeneous medium such as the Milky Way ISM, SNe carry shocked matter radially outward following the path of least resistance. As a consequence, the cool dense phase traced by CO receives less energy injection than the warm diffuse phase traced by \halpha, resulting in lower levels of turbulence in the corresponding VSF. Such behavior is also seen in hydrodynamical simulations of single SN remnants \citep[e.g.,][]{Martizzi15,2015ApJ...814....4L}. 

In Perseus and Taurus, the VSFs traced by \halpha~and CO show a good agreement in both the amplitudes and slopes in these regions, except for the range $\ell \lesssim 7$~pc in Perseus, which suffers from projection effect due to the thickness of the cloud \citep[see Section 4.2 of][for a detailed discussion]{Qi15}. At larger $\ell$, the VSF traced by \halpha~and stars also closely follow each other. Such trend is consistent with the evolutionary history of these regions with no clear imprint of recent SN explosions. Without this important local driver of turbulence, different phases of gas are well-coupled, and the young stars formed out of these molecular clouds also largely retain the same memory of turbulence. 

The difference of \halpha~VSFs in different regions show that ISM turbulence can be unevenly distributed spatially as a result of local drivers. Turbulence can also be unevenly distributed between different phases of the ISM (e.g., Orion and Ophiuchus). Section~\ref{sec:halpha_discussion} and Figure~\ref{fig:emission} further demonstrate that even with the same gas tracer \halpha, the level of turbulence can vary depending on its brightness. In addition, we note that local energy injection from supernova explosions is also intermittent temporally. Using the \halpha~VSF in Ophiuchus, we estimate the eddy turnover time at the driving scale to be:
\begin{equation}
\frac{\ell}{v}\sim \frac{20-30\, \rm{pc}}{5\, \rm{km/s}}\sim 4-6\, \rm{Myr}.
\end{equation}
The \halpha~VSF in Orion gives a similar estimate. Typically, it takes a few eddy turnover times to establish a steady-state turbulent flow. On the other hand, given the size of these regions, the average time interval between supernovae is on the order of a few Myr \citep{1990ApJ...352..222N, 2015ApJ...814....4L}. Each event also injects energy as a pulse with a very short duration. Therefore, both the ``on'' and ``off'' states of the driver are short compared with the timescale required to achieve a steady state. Thus ISM turbulence in molecular complexes is often not in a steady state.

\section{Conclusions} \label{sec:conclusions}

In this work, we analyze the kinematics of young stars and gas in four nearby star-forming regions: Orion, Ophiuchus, Perseus, and Taurus. For each region, we compute the first-order VSF of young stars and \halpha~gas, and compare the results against each other, as well as analysis of CO molecular cores from the literature.

We find that the VSFs traced by young stars in all four regions exhibit similar properties largely in agreement with Larson's law and reveal a universal scaling of turbulence in the ISM. This also confirms our previous finding in \citet[][]{Ha21} that young stars retain the memory of turbulence in their natal molecular clouds, and thus can be used to measure the turbulent kinematics of molecular clouds. 

The VSFs of \halpha~are more diverse, likely as a result of local energy injection. In regions with recent SN activities such as Orion and Ophiuchus, the VSFs of \halpha~are higher in amplitude compared to those without. The \halpha~VSFs are also higher than the VSFs of stars and CO in the these regions, while regions without recent SN explosions (Perseus and Taurus) show well-coupled turbulence between different phases. Within each individual region, the high-emission \halpha~tends to show higher VSFs than the low-emission \halpha.

Our findings support a complex picture of the Milky Way ISM, where turbulence can be driven at difference scales and inject energy unevenly into different phases. Future observations with more tracers of turbulence and on more star forming regions will help us better understand the diversity and complexity of ISM turbulence. In addition, high-resolution numerical studies can help further our understanding of the coupling between different phases and relevant timescales.

\section*{Acknowledgments}
We thank Shmuel Bialy, John Forbes, Di Li, Mordecai-Mark Mac Low, Lei Qian, Amiel Sternberg, and Yuan-Sen Ting for the helpful discussions. We thank Zhijie Qu for sharing the Galactic rotation model.
H.L. and S.X. are supported by NASA through the NASA Hubble Fellowship grant HST-HF2-51438.001-A and HST-HF2-51473.001-A, respectively, awarded by the Space Telescope Science Institute, which is operated by the Association of Universities for Research in Astronomy, Incorporated, under NASA contract NAS5-26555. 
Y.L. acknowledges financial support from NSF grant AST-2107735 and the College of Science and College of Engineering at UNT. Y.Z. is supported by NASA through a grant (HST-AR-16640.001-A) from the Space Telescope Science Institute, which is operated by the Association of Universities for Research in Astronomy, Incorporated, under NASA contract NAS5-26555. The Wisconsin \halpha~Mapper and its \halpha~Sky Survey have been funded primarily by the National Science Foundation. The facility was designed and built with the help of the University of Wisconsin Graduate School, Physical Sciences Lab, and Space Astronomy Lab. NOAO staff at Kitt Peak and Cerro Tololo provided on-site support for its remote operation.


\vspace{5mm}
\facilities{Gaia, Sloan, WHAM}

\software{astropy \citep{2013A&A...558A..33A,2018AJ....156..123A}, matplotlib \citep{matplotlib}, numpy \citep{numpy1,numpy2}, pandas \citep{pandas1,pandas2}, scipy \citep{scipy}, Sagitta \citep{2021AJ....162..282M}}

\bibliography{ha}{}
\bibliographystyle{aasjournal}

\appendix
\renewcommand\thefigure{\thesection.\arabic{figure}}
\renewcommand\thetable{\thesection.\arabic{table}}

\twocolumngrid
\setcounter{figure}{0}

\section{Selection of the \rophi~dark cloud within the Upper Sco region}
\label{sec:Upper_Sco}

\begin{figure}
    \centering
    \includegraphics[width=\columnwidth]{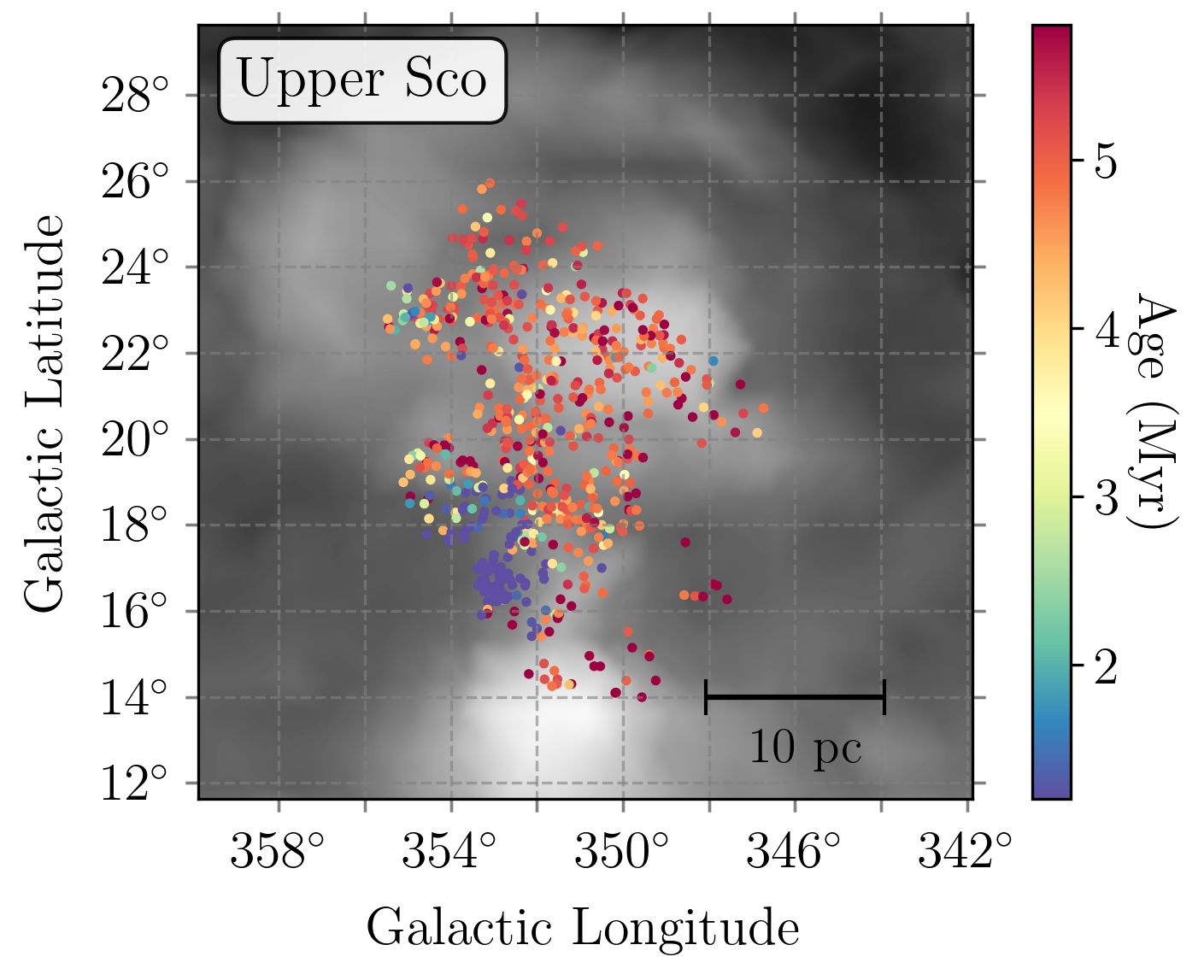}
    \includegraphics[width=\columnwidth]{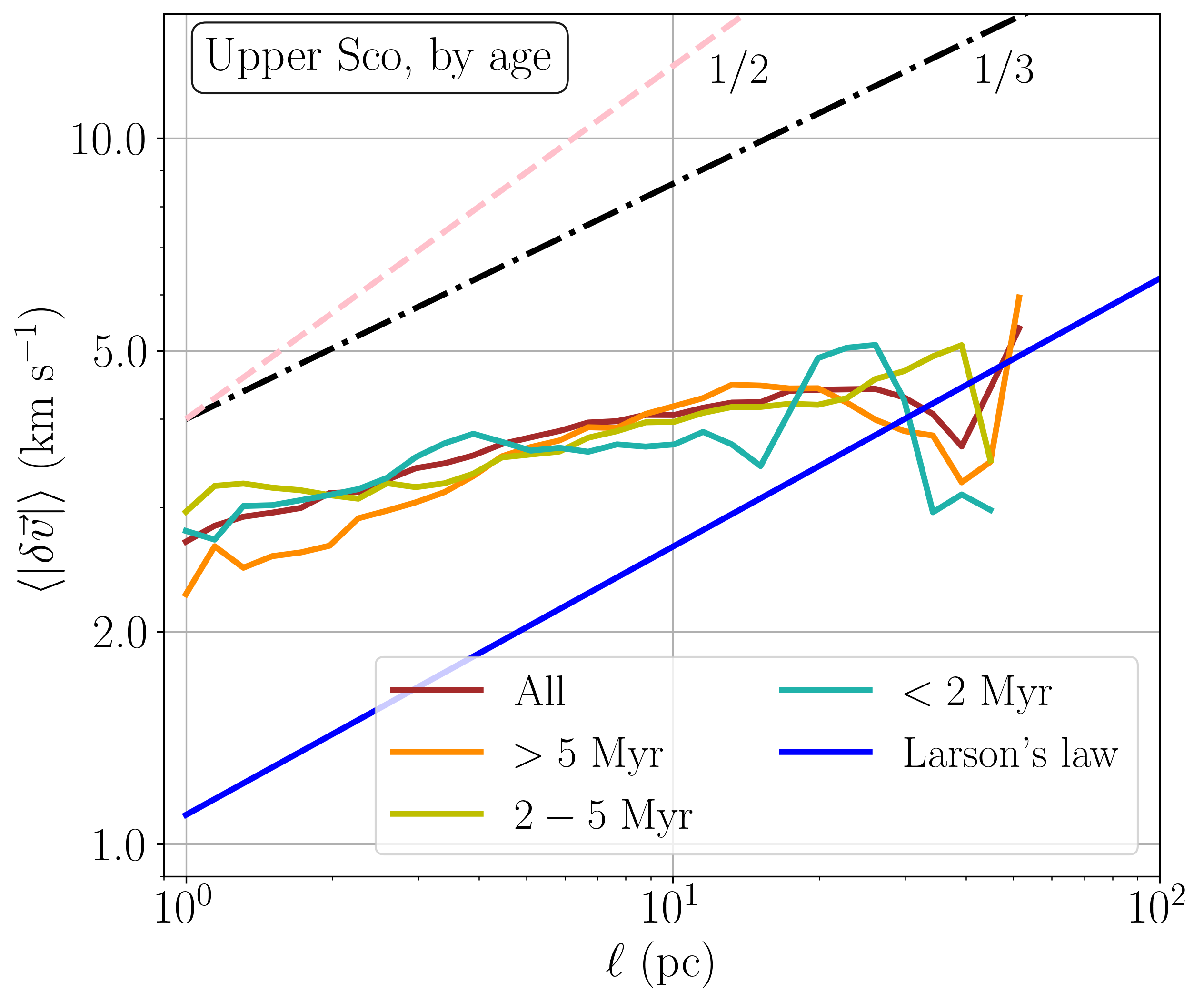}
    \caption{Top panel: positions of young stars color-coded by their age. Stellar age is estimated using the Sagitta package developed by \citet{2021AJ....162..282M}. Background shows \halpha~flux from the WHAM survey. Bottom panel: the VSF of stars in Upper Sco within three age brackets. Only stars younger than 2 Myr show characteristic of SN energy injection.}
    \label{fig:uppersco_age}
\end{figure}
In the main body of this paper, we select only stars within the actively star-forming region Ophiuchus. Here, we show the analysis results when selecting the entire Upper Sco region and discuss how the VSF of the stars depend on the stellar age.

The top panel of Figure \ref{fig:uppersco_age}~shows the spatial distribution of stars in Upper Sco and their age. Most of the stars outside Ophiuchus~are older than 4 Myr, while stars inside Ophiuchus~are much younger, primarily younger than 2 Myr. The SN explosion in this region traced by runaway stars took place $\sim 2$~Myr ago \citep[][]{Ophi_SN20}, so the majority of stars in Upper Sco did not receive energy injection from this event, and thus would not exhibit characteristics of such energy injection in their VSF. Indeed, we see this trend in the lower panel of Figure \ref{fig:uppersco_age}. With the entire Upper Sco region taken into account, the VSF shows only a slight bump at $\sim 25$~pc, while both the VSF of stars older than 5 Myr and of stars between 2 and 5 Myr do not show any discernible energy injection scale. Meanwhile, the VSF of stars younger than 2 Myr shows a clear energy injection scale at $\sim 25$~pc, similar to what we obtain in the main paper. Because of these results, we elect to use only stars categorized as members of Ophiuchus a majority of which are younger than 1 Myr, to analyze in the main body of this paper.

\section{Effect of Galactic Rotation on the VSF of \halpha}
\label{sec:rotation}

\begin{figure}
    \centering
    \includegraphics[width=\columnwidth]{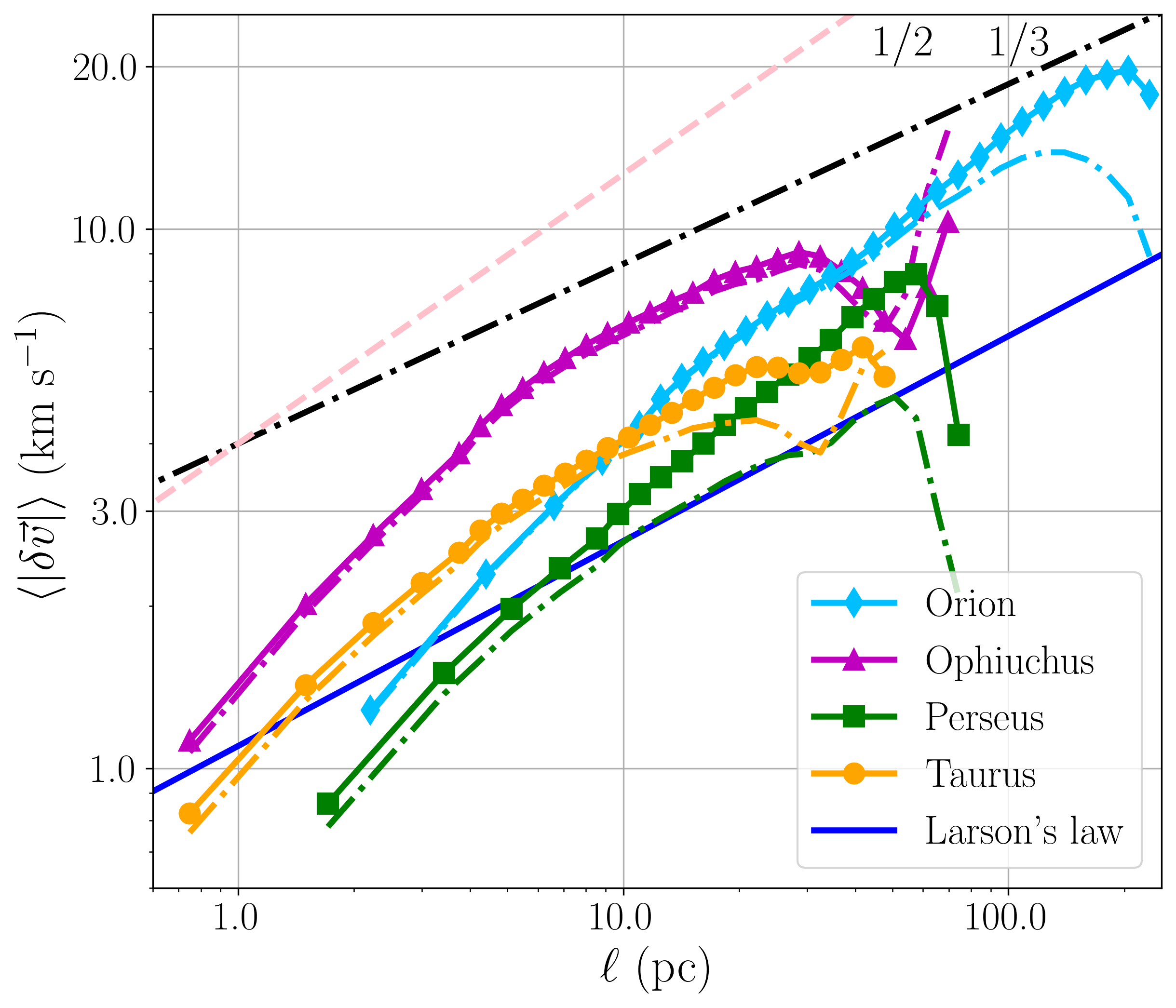}
    \caption{The VSFs of \halpha, with (solid lines) and without (dotted-dashed lines) a correction to the Galactic rotation \citep{Qu20}.}
    \label{fig:rotation}
\end{figure}

\begin{figure*}[ht]
    \centering
    \includegraphics[width=0.5\linewidth]{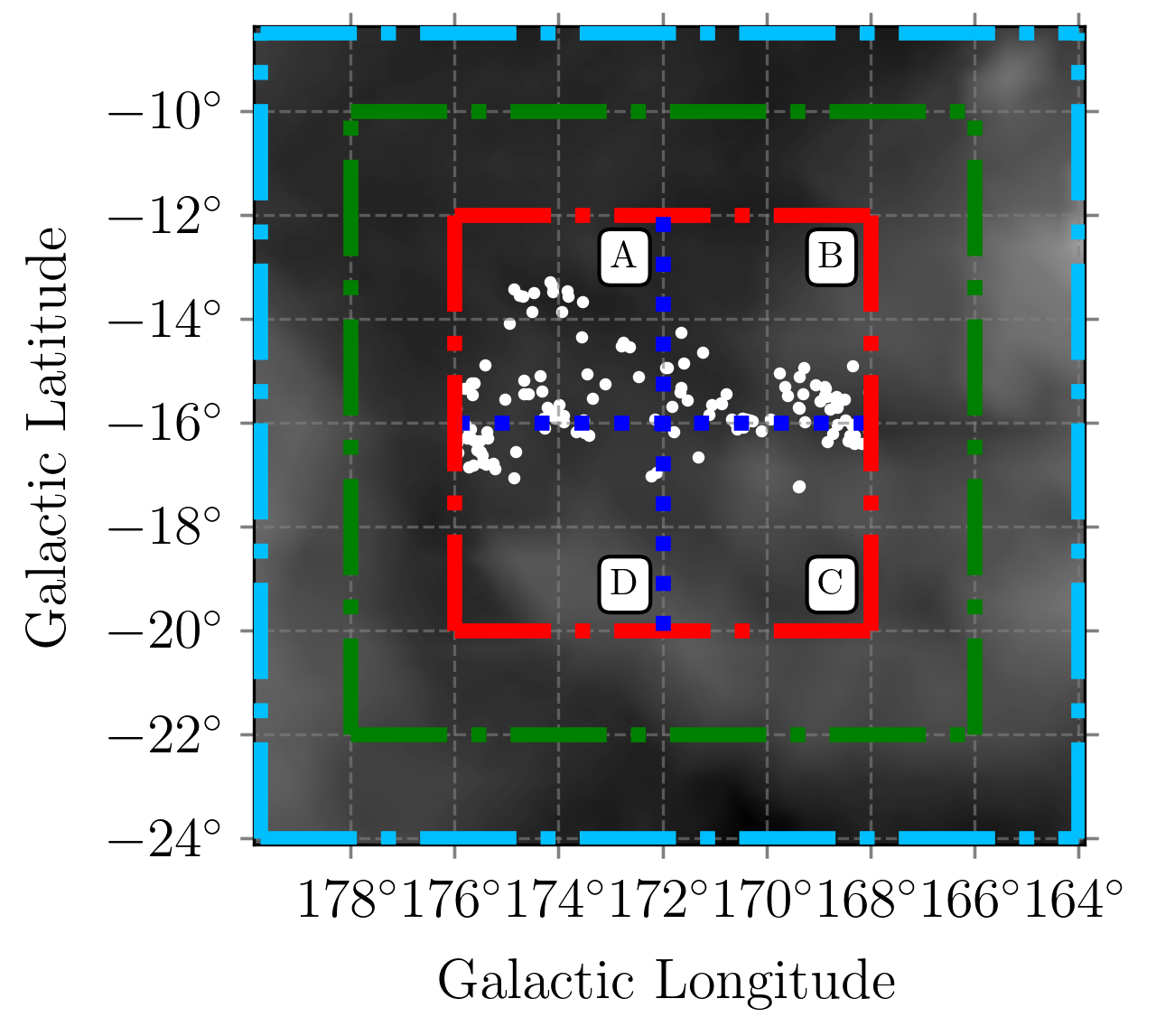} \\
    \includegraphics[width=0.45\linewidth]{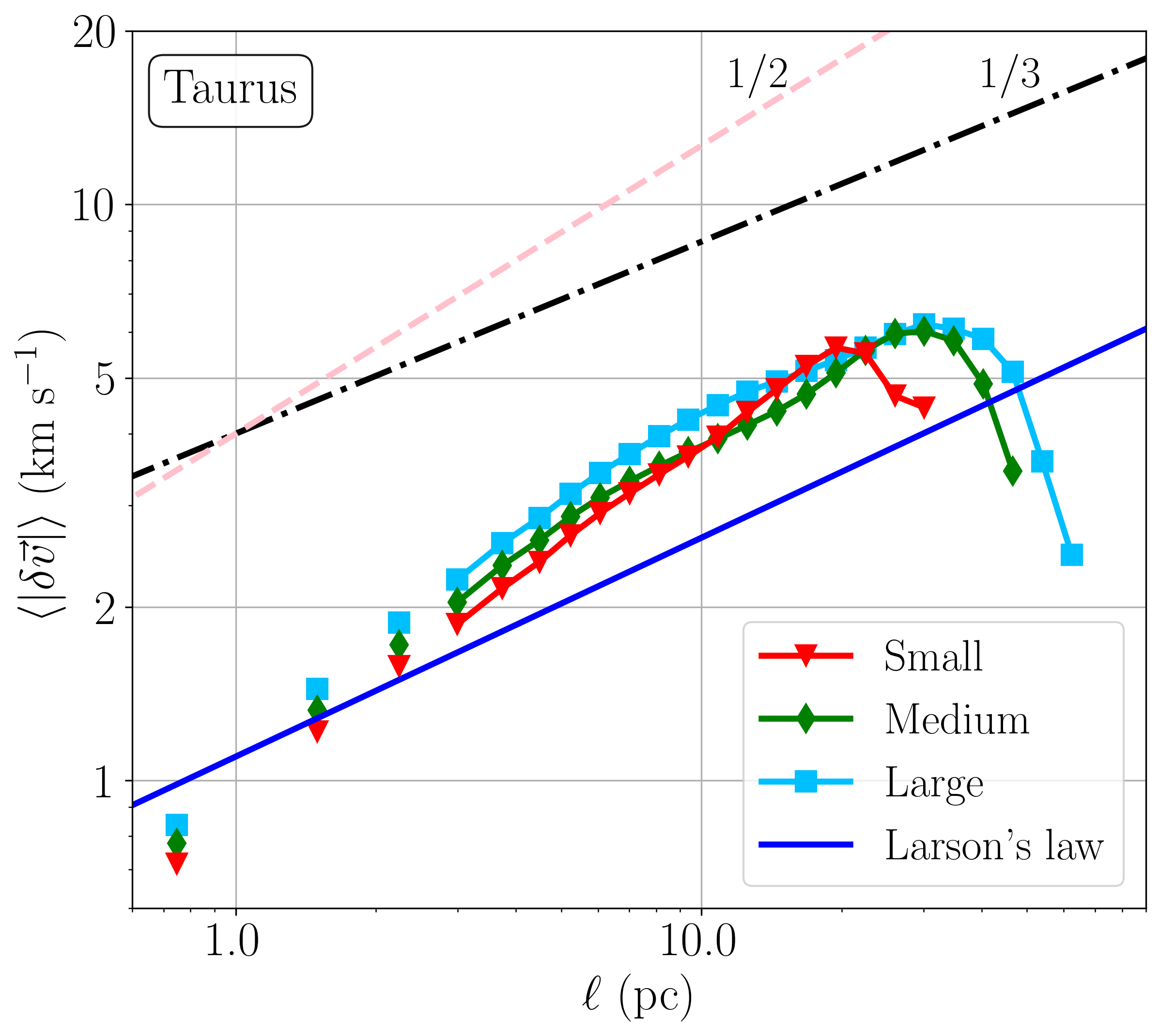}
    \includegraphics[width=0.46\linewidth]{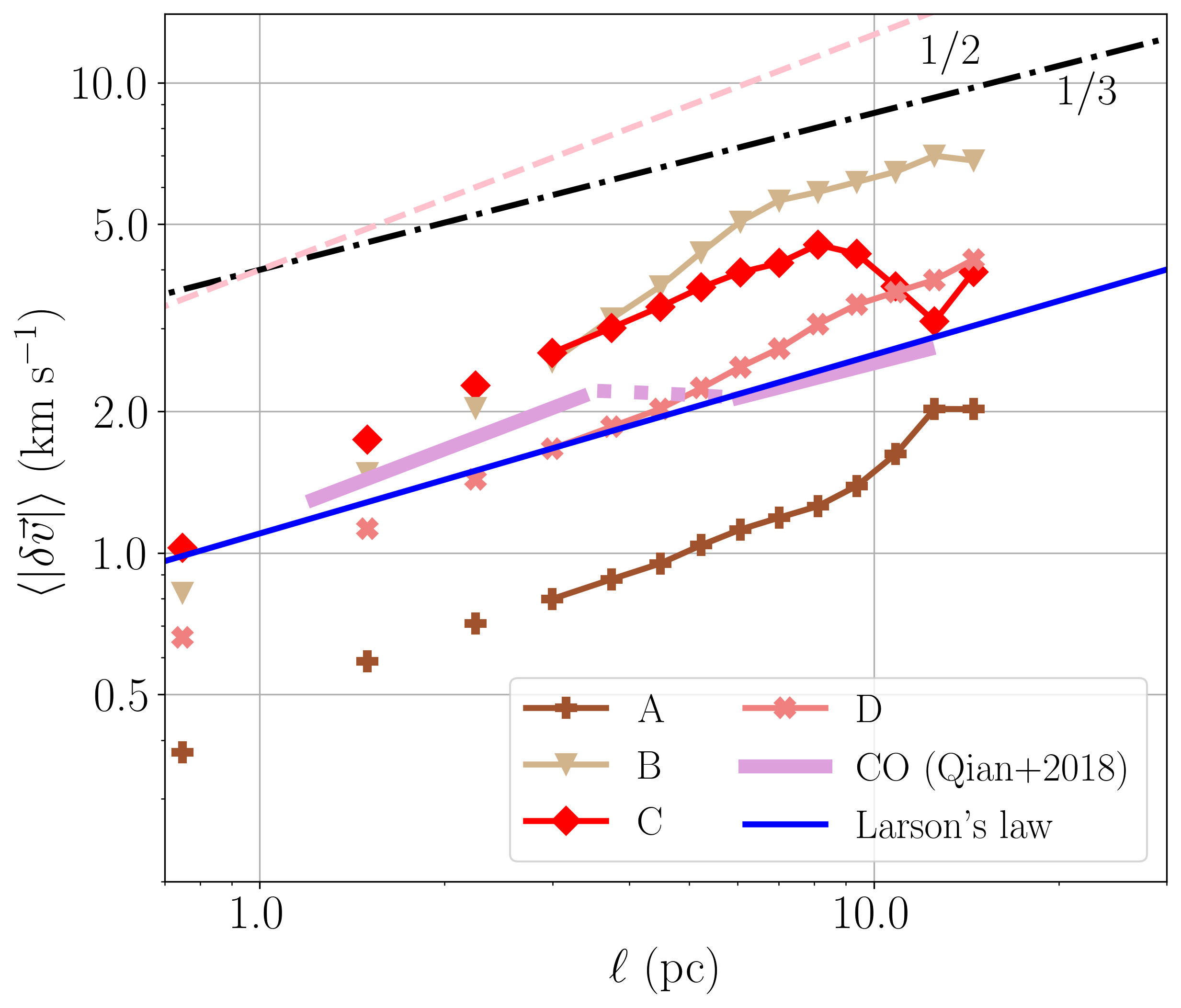}
    \caption{Top panel: illustration of our \halpha~window tests. The red box is the smallest window within which all stars in Taurus belong. The two boxes in green (the window used in the main analysis) and light-blue are each 4 deg$^2$~bigger than the former. Inside the red box, we also test the variation of the VSF within the region by dividing the window into four equal-sized boxes. Bottom left panel: The VSF of Taurus traced by \halpha, plotted across three different window sizes.
    Bottom right panel: The \halpha~VSFs of Taurus in the four quadrants within the red window. }
    \label{fig:aperturesize}
\end{figure*}

Figure \ref{fig:rotation} shows the VSF traced by \halpha~in four regions with and without Galactic rotation correction. In three out of the four regions, when a rotation profile is applied, the VSF does not significantly differ from the VSF without a rotation correction up to $\sim 20$~pc. This behavior is expected, since at small $\ell$ the rotation correction between two points are very similar. Thus, $\delta \vec{v}$ between two points essentially cancel out the rotation correction, and the shape of the VSF is preserved. For two points separated by a large distance, the rotation correction applied to them can differ more, and hence results in a shift in the VSF, as Figure \ref{fig:rotation} shows. However, the overall shifts in the VSFs due to rotation correction are moderate and do not affect the conclusions of our work.

\section{Effect of the size and location of the \halpha~window}
\label{sec:windowsize}

Another potential source of bias comes from the exact size and location of the window chosen to calculate the VSF of \halpha~within each region. In Section \ref{subset:eachregion_result}, we caution that the bumps seen in the VSF of \halpha~is a potential artifact since it is close to the size of the window. Such effect is noted as the apparent driving scale in simulations of multi-phase turbulence, and is equal to 1/2 of the box size \citep[e.g.,][]{2022MNRAS.510.2327M}. To test whether this effect also presents in the VSF of \halpha, we select a solid angle that encompasses all the stars in a region, and compute the VSF within this window. We then compute the VSF for a window that is 4 deg$^2$ larger, and repeat this a second time (see top panel in Figure~\ref{fig:aperturesize}). 

In all four regions, the VSFs of \halpha~are independent of the window size up to $\ell \sim L/2$, where $L$ is the window size. In Taurus and Perseus, the bump in the VSF shifts to larger $\ell$ when the window size increases. Thus we conclude that in these regions, the VSF bump is likely caused by the box limit rather than a real physical driving scale. As an example, the bottom left panel of Figure \ref{fig:aperturesize} shows the VSF of \halpha~in Taurus as a function of window size.

We also use Taurus as an example to illustrate the spatial variation of turbulence on small scales. We divide the smallest box (8 deg$^2$) in the previous experiment into four equal-sized quadrants. The bottom right panel of Figure \ref{fig:aperturesize} shows the \halpha~ VSFs of these four quadrants in Taurus. Although the slopes of these four VSFs are roughly consistent with one another, their amplitudes vary by up to a factor of three, indicating potentially large variations in turbulence between local patches of the ISM. The VSF of CO is in the best agreement with the bottom-left window ($172^{\circ} < l < 176^{\circ}$~and $-20^{\circ} < b < -16^{\circ}$), where many CO cores were also identified \citep[see Figure 2 of][for detailed locations of CO cores in Taurus]{Qi12}.

\end{document}